\documentclass[psfig]{aa}
\usepackage{psfig}
\begin{document}
\thesaurus {11.01.2, 11.07.1, 11.09.1 NGC 3783, 11.11.1, 11.16.2, 11.19.2}

\title{HI Spatial Distribution in the Galaxy NGC~3783}

\author{  J.A. Garc\'{\i}a-Barreto\inst{1}, F. Combes\inst{2},
B. Koribalski\inst{3}, J. Franco\inst{1} }

\institute{Instituto de Astronom\'{\i}a, Universidad Nacional Aut\'onoma 
       de M\'exico, Apartado Postal 70-264, M\'exico D.F. 04510, M\'exico
\and
  DEMIRM, Observatoire de Paris, 61 Av. de l'Observatoire,     
 F-75014 Paris, France
\and
  Australia Telescope National Facility, CSIRO, P.O. Box 76, 
       Epping, NSW 2121, Australia}

\offprints{J.A. Garc\'{\i}a-Barreto \hfill\break(e-mail: tony@astroscu.unam.mx)}   

\date{Received April 1999; accepted June 1999}

\maketitle
\markboth{Garc\'ia-Barreto et al.}{HI Spatial Distribution in the Galaxy NGC~3783}

\begin{abstract}
	We have mapped the emission from atomic hydrogen at $\lambda$=21 cm
from the galaxy NGC~3783 with the Australia Telescope Compact Array. 
Our main results are: 
{\bf a)} the HI morphology is irregular and perturbed, gathered in three blobs
apparently unrelated to the optical morphology;
{\bf b)} the observed HI velocity distribution
indicates a normal disk in differential rotation with a constant velocity out
to a radius of 160$''$ (30 kpc), 
{\bf c)} the inclination of the disk is about 
25$^{\circ}$ with the kinematic major axis at a position angle slightly 
different from that of the stellar bar, 
{\bf d)} the HI mass inside a radius of
18$''$ is only $2.1\times10^7$ M$_{\odot}$, the total HI mass within 180$''$
is $1.1\times10^9$ M$_{\odot}$ and the dynamical mass is $2\times10^{11}$ 
M$_{\odot}$. The bulk of the gas in NGC~3783 is outside the diameter of the
stellar bar;
{\bf e)} Numerical simulations of the gas flow in the barred potential derived
from the red image indicate that the pattern speed is $\Omega_p$ = 38 km/s/kpc:
the ring of H$\alpha$ emitting regions encircling the bar would then correspond to UHR,
and the H$\alpha$ accumulation in the center to a nuclear ring.
Various possibilities are discussed to account for the active nucleus fuelling.


\end{abstract}

\section{Introduction}

	The activity observed in the nucleus of Seyfert 1 galaxies requires 
gas fuelling, either constant or episodic, of a few tenths to a few M$_{\odot}$ 
yr$^{-1}$ for a significant period of time. Some authors have suggested as a
likely possibility that fuelling may be a direct consequence of a stellar
bar (Simkin, et al.  1980; Shlosman et al. 1989) which causes
radial motions of gas in the disk of the surrounding spiral galaxy. Therefore it is
important to determine the total amount of neutral hydrogen (HI) gas, its
spatial distribution and the rotation curve. 

	NGC~3783, a Seyfert\,1 SBa galaxy, shows a broad H$\alpha$ line and 
lines of high ionization atoms (Pelat et al. 1981; Atwood et al.
 1982; Evans 1988; Winge et al. 1992, George et al.
1995). The optical structure of NGC~3783 has a diameter of $1'.9\times1'.7$ 
at an inclination angle of $i = 23^{\circ}$ with a bright compact nucleus, a 
stellar bar at a position angle of $PA_{\rm b} \approx 163^{\circ}$ and radius
$r_{\rm b} = 18''$ (Mulchaey et al. 1997), a bright inner ring just
outside of the stellar bar and small pitch angle spiral arms (Kennicutt 1981).

\begin{figure}
\psfig{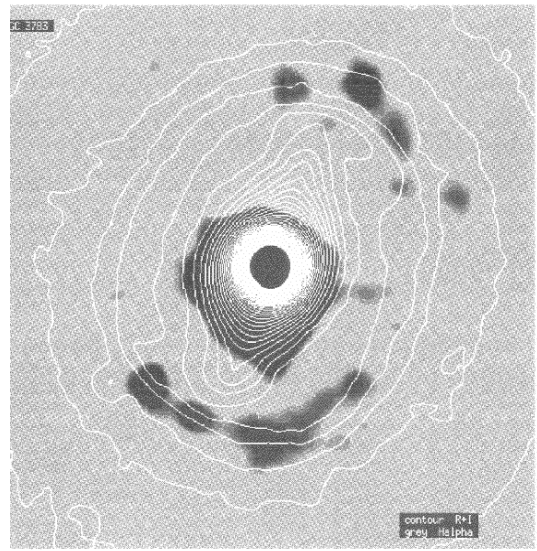}
\caption[8750.f1]{
Optical emission of the innermost continuum light of NGC 3783 in the R plus I 
broadband filters (in contours) superimposed on the continuum-free H$\alpha$+[NII] 
emission (in grey scale) from Garc\'{\i}a-Barreto et al. (1996). North is up, East 
is to the left. Angular scale is such that distance from peak to peak H$\alpha$ 
emission outside the bar from the south east to the north west is about 37$''$.}
\label{opt}
\end{figure}

The peak to peak distances from the nucleus to the HII regions to the NW and SE are 
$\sim21''$ and $\sim16''$, respectively, (see Fig. \ref{opt}). 
H$\alpha$ emission has been detected from the innermost central region, from
regions just at the end of the stellar bar and from regions 
perpendicular to the bar (Forte et al. 1987; Winge et al. 1992; Garc\'{\i}a-Barreto et al. 1996 
and Fig. \ref{opt}). The nucleus of NGC 3783 is a strong X-ray emitter in the 2--10 keV band 
(Piccinotti et al. 1982) and in the 
0.2--3.5 keV band  (Fabbiano et al. 1992). The short time scale variability of 
its optical, X-ray and UV continua suggests a collimation outflow from NGC~3783 (Reichert et 
al. 1994; Stirpe et al. 1994; Alloin et al. 1995). Radio continuum emission has been detected 
from an unresolved central source at different frequencies indicating an optically thin 
synchrotron emission (Ulvestad \& Wilson 1984; Alloin et al. 1995). Its far infrared (FIR) 
IRAS fluxes are $f_{60\mu} \approx 3.37$~Jy and $f_{100\mu} \approx 5.12$~Jy with a 
two color dust temperature of $T_{\rm d} \approx 39$~K. In this paper we adopt a 
distance to NGC~3783 as 38.5 Mpc, $H_0$ = 75 km\,s$^{-1}$ Mpc$^{-1}$, as given by 
Tully (1988). The FIR luminosity is $L_{\rm FIR} \approx 3.1 \times 10^{43}$~ergs s$^{-1}$ 
(according to the formula by Helou et al. 1985). Spectroscopic 
studies reveal extremely broad Balmer lines with a full width at zero intensity (FWZI) 
of the H${\alpha}$ line of $\approx7,300\rightarrow10,000$ km\,s$^{-1}$ 
(Pelat et al. 1981; Evans 1988). Broad line components are
increasingly blue-shifted relative to the systemic velocity (Evans 1988).
The density, ionization parameter, and velocity dispersion of the emitting
clouds increase toward the central ionizing source
(Pelat et al. 1981; Atwood et al. 1982; Evans 1988;
Winge et al. 1992). The optical systemic velocity using lines of OI, OII and
NII is $v_{\rm sys}^{\rm opt} \approx 2930$ km\,s$^{-1}$ (Pelat et al. 
 1981), while the average velocity of the narrow line region is 2890 
km\,s$^{-1}$.
 
	Single dish observations of NGC~3783 indicate a total HI mass of 
$M_{\rm HI} = 3.6 \times 10^9$~M$_{\odot}$, and a systemic velocity of
$v_{\rm sys} \approx 2902$ km\,s$^{-1}$ (Huchtmeier \& Richter 1989). A
previous interferometric study of the dynamics and spatial distribution of
neutral hydrogen in NGC~3783 indicated that the total HI has an extent  similar 
to the weak surface brightness optical contours (25$\rightarrow28$ 
mag\,arcsec$^{-2}$), and that the velocity field exhibits irregular structures
(Simkin \& van Gorkom 1984). 

The present study focuses on the spatial distribution of the HI gas in NGC~3783
obtained with the Australia Telescope Compact Array (ATCA), giving an angular
resolution of 30$''$ and a spectral resolution of 6.7~km\,s$^{-1}$. In 
Section~2 we describe the HI observations. In Section~3 we describe the HI 
spatial distribution and velocity field, discuss possible physical scenarios
for the central gas kinematics in NGC~3783, and describe the central radio 
continuum emission. The dynamics of the central parts is investigated
through numerical simulations fitted to the observations in Section~4.
Our conclusions are summarized in Section~5.
	
\section{Observations and Data Reduction}
The observations of NGC~3783 were carried out with the 1.5A configuration of
ATCA in Narrabri, NSW,
on 1996 October 18, in the 21-cm neutral hydrogen line and the 13-cm radio 
continuum. The two intermediate frequency (IF) bands were centered at 1407 
and 2378 MHz with a total bandwidth of 8 and 128 MHz, respectively. The total
time on source was $\sim$9 hours. The pointing center was 
$\alpha$(J2000) = $11^{\rm h}\,39^{\rm m}\,01^{\rm s}.8$ and 
$\delta$(J2000) = $-37^{\circ}\,45'\,20''$, which is about 1$'$ away from 
the center of NGC~3783.

The initial calibration, data editing and final analysis were 
performed using the AIPS package. 
For both frequencies PKS\,1934--638 (14.893~Jy / 11.553~Jy) was used as flux
calibrator, PKS\,1151--348 (6.0~Jy / 4.6~Jy) and PKS\,1215--457 (4.8~Jy / 
3.6~Jy) as phase calibrators, and both PKS\,1934--638 and PKS\,0823--500
(5.8~Jy / 5.8~Jy) as bandpass calibrators.

\subsection{HI Observations}
The bandwidth of the first IF was divided into 512 channels spaced by 
15.625 kHz (or 3.37 km\,s$^{-1}$) and covered a velocity range from about
2000 to 3600~km\,s$^{-1}$. The line-free channels were averaged and 
subtracted using the AIPS task UVLSF.

\begin{figure}
\psfig{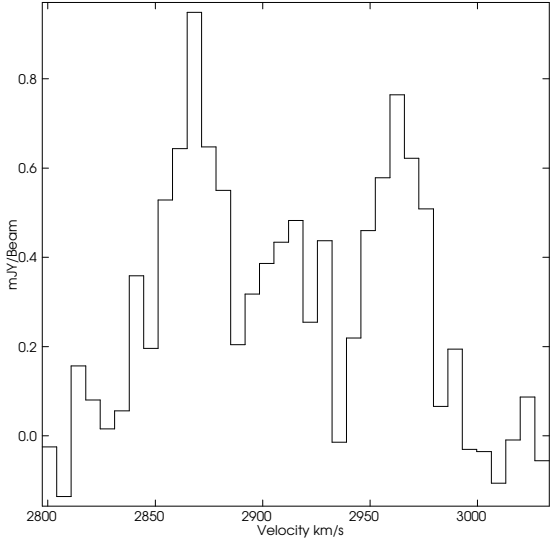}
\caption[8750.f2]{
HI spectrum of NGC 3783. Only 23 of the 35 channels show emission in the 
velocity interval from 2840 to 2990 km s$^{-1}$.}
\label{hispec} 
\end{figure}

The resulting line data were Fourier transformed using `natural weighting' to
produce a cube with dimensions 256 pixel $\times$ 256 pixel $\times$ 35 
channels with $8''$ pixels and a final channel separation of 6.74 km\,s$^{-1}$.
The data showed emission in the velocity range from $\sim$2840 to 2990 
km\,s$^{-1}$.
Each of the 35 channels was CLEANed and restored with a clean beam of 45$''$ (FWHM) in 
order to detect extended emission. The rms noise per channel was 
$\sim$1.7 mJy\,beam$^{-1}$. Only 23 out of the 35 cleaned channels show HI 
emission (see Fig.~\ref{hispec}). The integrated neutral hydrogen distribution and the 
corresponding velocity field were derived from the data by moment analysis 
using the AIPS task MOMNT. We smoothed the data both in velocity and space 
using Hanning (3 channels) and Boxcar (7 pixels) functions, respectively. 
Only fluxes greater than 2 mJy\,beam$^{-1}$ were included in the integration. 

\subsection{Radio Continuum Observations}
The narrow-band 20-cm continuum data were Fourier transformed using `uniform
weighting' to produce a cube with dimensions 1024 pixel $\times$ 1024 pixel
with 2$''$ pixels. It was CLEANed and restored with a clean beam of 
13\farcs5 (FWHM). The rms noise was about 0.12 mJy\,beam$^{-1}$.
The bandwidth of the second IF was divided into 32 channels; the central 
26 of those were averaged to form the 13-cm radio continuum image.
The data were Fourier transformed using `uniform weighting' to 
produce a cube with dimensions 1024 pixel $\times$ 1024 pixel with $2''$ pixels.
It was CLEANed and restored with a clean beam of  (FWHM).
The rms noise was 0.06 mJy\,beam$^{-1}$.

\section{Results and Discussion}

\subsection{HI Emission}

\begin{figure}
\psfig{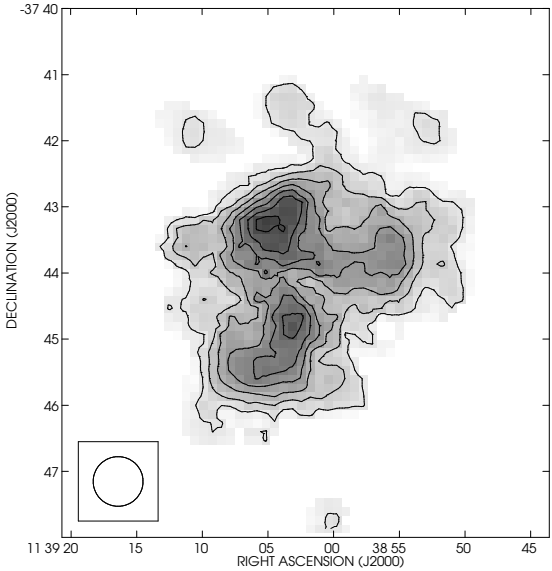}
\caption[8750.f3]{
Integrated neutral hydrogen distribution (zero moment) of NGC~3783. 
The contour levels are 0.05, 0.1, 0.15, 0.2, 0.25, 0.3, and 0.35 
Jy\,beam$^{-1}$ km\,s$^{-1}$. The first contour corresponds to an
HI column density of $6.25 \times 10^{19}$ atoms\,cm$^{-2}$.
The synthesised beam (45$''$) is indicated in the lower left corner.
North is to the top, east to the left.} 
\label{hi} 
\end{figure}

\begin{figure}
\psfig{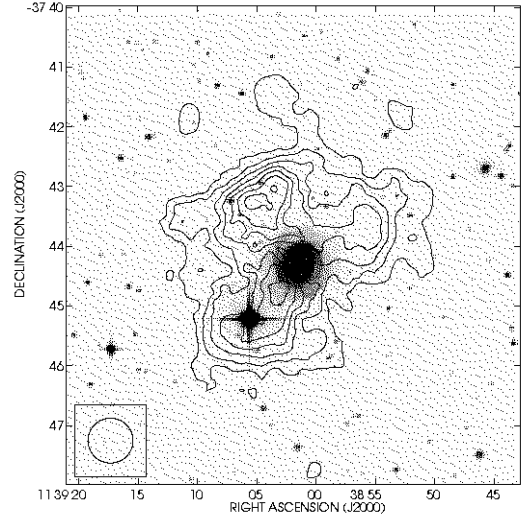}
\caption[8750.f4]{
Integrated neutral hydrogen 
distribution of NGC 3783 (in contours) superimposed on the innermost 
continuum light emission from a POSS plate (in grey scale). The optical 
emission at low levels (~25 mag arcsec$^{-2}$) extends to about 60$''$
to the south-west and about 120$''$ elsewhere.}
\label{hiopt} 
\end{figure}

\begin{figure*}
\psfig{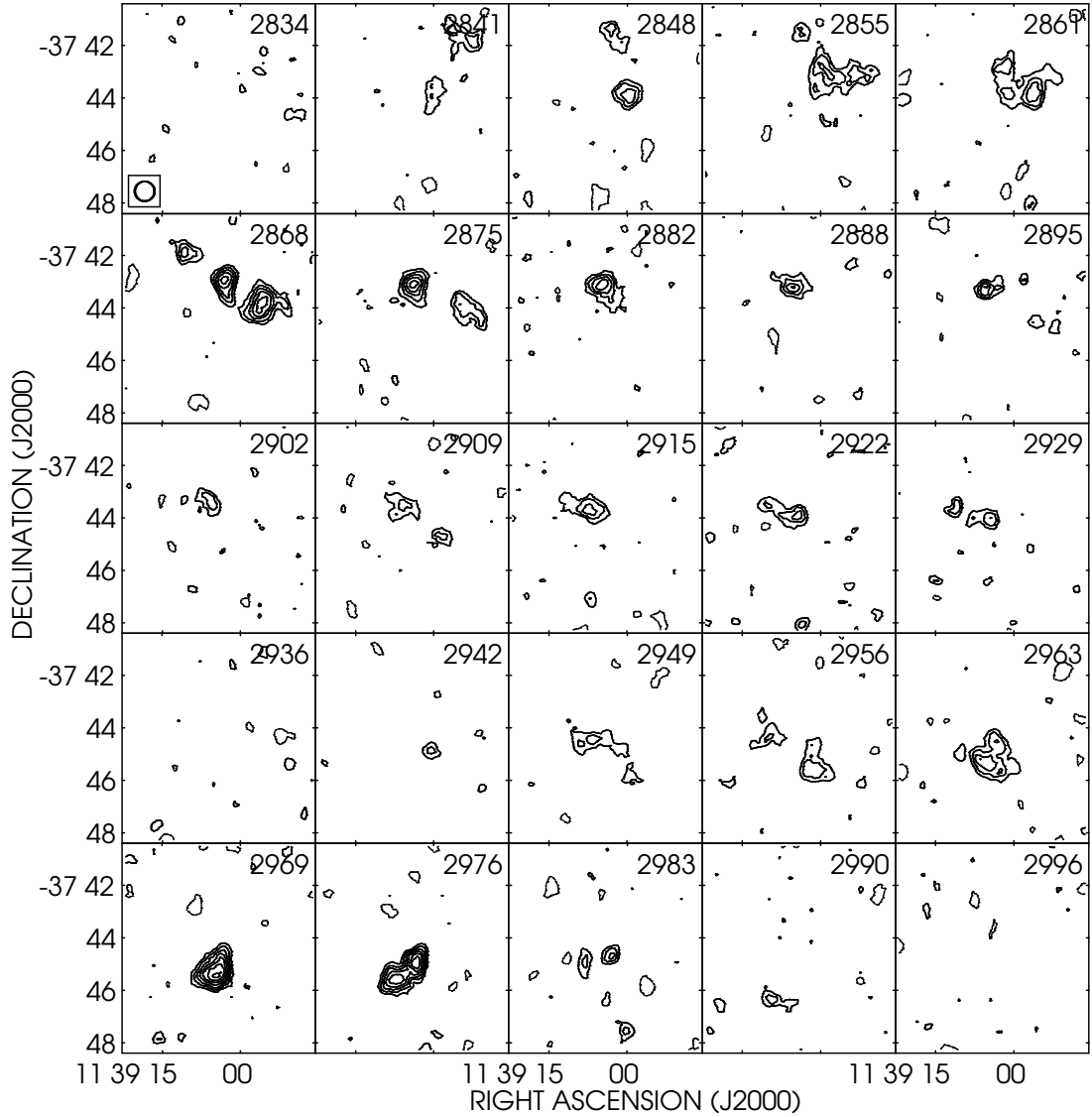}
\caption[8750.f5]{
HI channel maps of  NGC~3783. Contour levels are --4, 4, 6, 8, 10, 12, 14
and 16 mJy\,beam$^{-1}$. The restoring beam is 45$''$.}
\label{hichan} 
\end{figure*}

	The spatial distribution of HI in NGC~3783 is shown 
in Fig.~\ref{hi} and \ref{hiopt}. 
The map results from integrating the column density over the velocity 
range with HI emission. Fig.~\ref{hichan} shows the spatial distribution of HI in 
individual channel maps. The overall spatial distribution of the HI emission is that 
of a disk with the blue-shifted velocity to the NW, and the red-shifted velocity to the SE.

	The integrated flux density inside a radius of 180$''$ (34 kpc) 
is $F_{\rm HI}$ = 3.2 Jy~km\,s$^{-1}$. This value is a factor of 3 lower than the
single dish value of 10.3 Jy~km\,s$^{-1}$ (Huchtmeier \& Richter 1989). This difference 
is most likely a result of missing short spacings and our column density threshold of 
$N_{\rm HI} = 6 \times 10^{19}$ atoms\,cm$^{-2}$. More extended emission could 
be present in the galaxy at much lower levels. From our observations, the total
HI mass inside a radius of 180$''$ is $M_{\rm HI} \approx 1.1 \times 10^9$
M$_{\odot}$. The HI mass inside the stellar bar radius of 18$''$ (3.4 kpc) is only 
$\approx 2.1 \times 10^7$ M$_{\odot}$, i.e. 2\% of the total mass. The HI integrated 
column density map indicates that most of the neutral atomic gas is distributed outside 
the stellar bar with large concentrations up to 120$''$ (22.5 kpc) away from the 
nucleus to the NE, NW and SE of the galaxy (see Fig.~\ref{hiopt}). The HI extent to the SW 
is only about $60''$ (11.2 kpc). The kinematical center was found to be at 
$\alpha$(J2000) = $11^{\rm h}\,39^{\rm m}\,01^{\rm s}.74$ and 
$\delta$(J2000) = $-37^{\circ}\,44'\,20\farcs2$. 

The rotation velocity, $v_{\rm rot}$, and the position angle, $PA$, of the 
line of nodes (major axis) were obtained by fitting the velocity field 
in inclined annuli using the AIPS program GAL. The kinematic major 
axis of the HI distribution is at $PA \approx 145\,\pm\,2^{\circ}$. This 
angle differs by about $15^{\circ}$ from the $PA$ of the bright stellar bar
which is at $\approx 163^{\circ}$ (Garc\'{\i}a-Barreto et al. 1996; Mulchaey, 
Regan \& Kundu 1997). The kinematic minor axis of NGC~3783 is thus at $PA
\approx 55^{\circ}$ indicating that the regions NE and SW of the galaxy are
at velocities near the systemic velocity of the galaxy.

\begin{figure}
\psfig{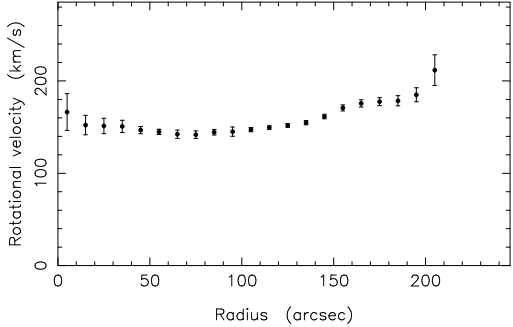}
\caption[8750.f6]{
Rotation velocity as a function of radius including both the receding 
and approaching sides of the galaxy.} 
\label{hirot} 
\end{figure}

\begin{figure}
\psfig{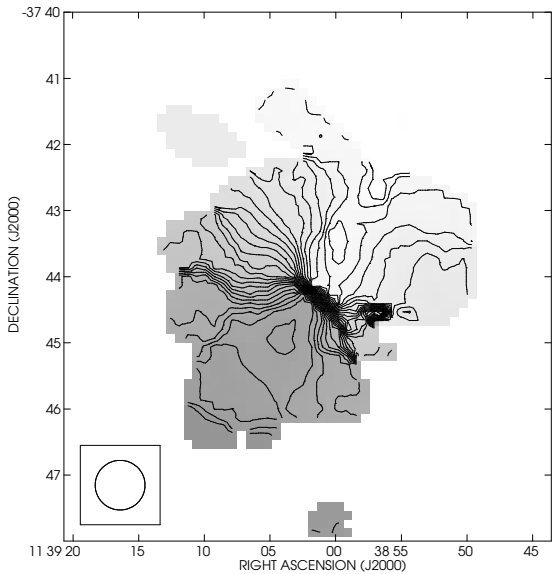}
\caption[8750.f7]{
 Intensity weighted mean velocity field (first moment) of NGC~3783.
   The contour levels go from 2845 to 2990, step 5 km\,s$^{-1}$.} 
\label{hivel} 
\end{figure}

Fig.~\ref{hirot} shows the observed rotation velocity, while Fig.~\ref{hivel} 
presents the velocity field. The fitted systemic velocity is $v_{\rm sys} \approx 
2910\,\pm\,5$ km\,s$^{-1}$. The observed velocities show an average 
rotational velocity of 160 km\,s$^{-1}$, with a maximum of about
170 km\,s$^{-1}$, and then a trend of decreasing velocity down to about 
130 km\,s$^{-1}$ at radii between 60$''$ to 70$''$. Then increasing 
velocities up to 160 km\,s$^{-1}$ at radii between 90$''$ to 100$''$, showing
a rotation curve similar to other spiral galaxies and our Galaxy (see 
Fig.~6 of Giovanelli \& Haynes 1988). The total dynamical mass inside 
a radius of 180$''$ is $M_{\rm dyn} \approx 2 \times 10^{11}$~M$_{\odot}$,  
for an inclination of $i = 25^{\circ}$ and a constant rotation velocity of 
160~km\,s$^{-1}$.

The spatial distribution of HI and its velocity field 
indicate gas rotating in a disk. Since there is little 
 HI in the stellar bar, and the optical image does not have bright 
spiral arms, it is difficult to assign any correspondence between 
the optical features (nucleus, bar, arms) and the HI distribution.

\par
\subsection{Origin of the gas in the central region}
\par

	The activity observed in the nucleus of NGC~3783 requires gas 
fuelling. One likely possibility for gas to be present in the inner few arcseconds is 
inflow from the disk. In terms of galactic dynamics, this could be achieved 
by any combination of processes leading to angular momentum exchange in the disk of 
the host galaxy (Lynden-Bell \& Kalnajs 1972; Lynden-Bell \& Pringle 1974; Phinney 1994). 
Possibilities are: inflow associated with the presence of a non-axisymmetric potential 
(e.g Athanassoula 1992; Friedli \& Benz 1993; Habe \& Wada 1993; Friedli \& Martinet 1997), 
inflow due to the perturbations exerted by ram pressure from intergalactic gas 
(Kritsuk 1983), inflow due to tidal or direct interactions with a nearby companion 
(e.g. Toomre 1978; Noguchi 1988; Salo 1991; Horellou \& Combes 1993), and 
inflow due to accretion of cooling flows (Fabbiano et al. 1989).  

	Considering first the possibility of a companion, there is only one 
candidate seen in the optical images. A small galaxy at $\alpha$(J2000) =
$11^{\rm h}\,38^{\rm m}\,45^{\rm s}.62$ and $\delta$(J2000) = 
$-37^{\circ}\,42'\,40\farcs9$ with an approximate apparent diameter of 
35$''$ lies $\sim3\farcm6$ to the NW of NGC~3783
(visible in Fig. \ref{hiopt}). We have not detected any HI at this
position, within our velocity range from 2000 to 3600 km/s.
Since nothing is known about this galaxy in terms of its redshift 
or distance, and since it does not show any apparent signs of tidal interaction 
with NGC~3783, we may assume that it is a background object.
Since the distribution of HI around NGC 3783 is very irregular, and
appears un-relaxed, a likely possibility then is that it has swallowed
recently an HI-rich companion; this would explain the nuclear activity.
However, we have no way to confirm this hypothesis.
In the absence of nearby large optical companions, 
or optical signs of tidal interactions, we will assume in the following 
that gas responds only to the action of a non-axisymmetric 
gravitational potential. The direction of the bar torques on the gas
depend strongly on the location of the resonances, and therefore on
the bar pattern speed (see e.g. Combes 1988). To infer this parameter
from observations, we have performed numerical simulations and they 
are presented in Section ~ 4.

\subsection{Radio Continuum Emission}
The radio continuum images show a central source in NGC~3783, at
$\alpha$(J2000) = $11^{\rm h}\,39^{\rm m}\,01^{\rm s}.63$ and
$\delta$(J2000) = $-37^{\circ}\,44'\,20\farcs.0$,
with peak flux densities of 27.8 mJy\,beam$^{-1}$ at 20-cm, 
19.5 mJy\,beam$^{-1}$ at 13-cm, and faint emission surrounding the nucleus. 
The total flux densities at 20- and 13-cm are about 36.0 and 30.4 mJy, 
respectively.

The radio continuum emission from the disk that is related 
with star formation and evolution suggests in NGC 3783, based in the weak 
H$\alpha$ fluxes in the NW and SE outside the stellar bar radio continuum fluxes 
of the order of few tenths of $\mu$Jy, well below the current detectability with
reasonable integration times.

The spectral index between 1407 MHz and 2378 MHz is $\alpha \approx -0.8$.
 Previous high
resolution continuum observations showed an unresolved source with a peak 
flux density of 32.7 mJy\,beam$^{-1}$ at 1.4 GHz (Alloin et al. 1995), 13 mJy\,beam$^{-1}$ at 4.8 GHz 
(Ulvestad \& Wilson 1984), 6.1 mJy\,beam$^{-1}$ at 
8.4 GHz, and  2.6 mJy\,beam$^{-1}$ at 14.9 GHz (Alloin et al. 1995).

Our continuuum maps confirm the non-existence of nuclear extended emission 
from NGC~3783 (e.g., radio lobes). In general, Sy\,1's are predominantly less 
luminous than Sy\,2's galaxies (de Bruyn \& Wilson 1978; Ulvestad \& Wilson
1984; Baum et al. 1993), the thermal contribution is negligible in the 
broadline region (Ulvestad et al. 1981), and the radio continuum
is mostly non-thermal (de Bruyn \& Wilson 1978). The detected radio continuum 
radiation from the central region coincides with the distribution of the central 
H$\alpha$ emission, but our observations cannot identify the physical 
origin of the emission: either star formation and its subsequent evolution in 
circumnuclear regions ( e.g. Terlevich et al. 1992, 1994; 
see review by Dultzin-Hacyan 1997) or activity from the nucleus of the galaxy. The 
spectral index of the radio continuum emission in NGC~3783 
indicates that the radiation is optically thin synchrotron emission. 
The power at 20 cm is only $P_{\rm 20cm} = 4.6 \times 10^{20}$
W\,Hz$^{-1}$.

\section {Numerical simulations}

 In order to gain more insight in the NGC 3783 galaxy dynamics,
we compute the response of gas test-particles 
in the potential created essentially by the stars, (e.g. Garc\'{\i}a-Burillo 
et al. 1994; Sempere et al. 1995). A realistic galaxy potential has been derived from
the red band image. We assume that the barred disk can be
fitted by a single and well-defined wave pattern, characterized by 
the pattern speed $\Omega_p$. Since the periodic orbits, and
the consequent gas behaviour, strongly depend on $\Omega_p$,
the observed morphology of the gaseous and stellar disks will
only be recovered with a narrow range of pattern speeds.
 The method has been shown to be very sensitive to this parameter
(e.g. Garc\'{\i}a-Burillo et al. 1993). 

\subsection { Method }

\begin{figure}
\psfig{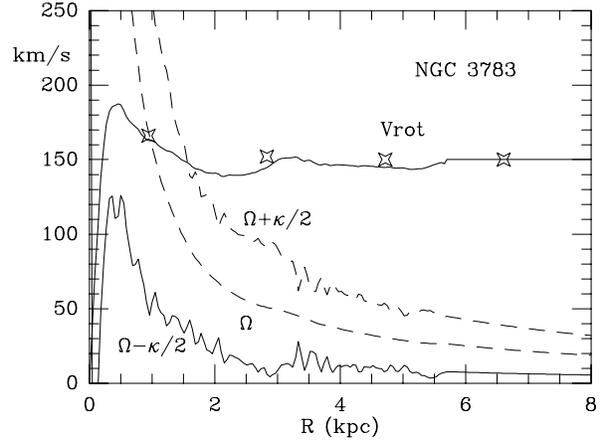}
\caption [8750.f8]{Rotation curve obtained from the galaxy potential, derived 
from the red image, and adopted for the simulations.
The stars are the data points derived from the HI observations. 
Also displayed are the derived principal resonance 
frequencies $\Omega$, $\Omega$-$\kappa$/2 
and $\Omega$+$\kappa$/2. } 
\label{vcir} 
\end{figure}

The gravitational potential is obtained from Fig. \ref{opt} in the 
following way. The image is first deprojected
using an inclination of 25$^\circ$, and a $PA$ of 145$^\circ$ (cf section 3).
The resulting image is then Fourier transformed on a 2D grid of 
256x256. 
The image used extends up to a maximum radius of 5.8kpc,
since the signal-to-noise decreases outwards. The spatial 
resolution is therefore of the order of the cell size, i.e. 45 pc.
Outside our maximum radius, the potential is extended analytically, in an
axisymmetric manner, and in order to have a flat rotation curve.
Fig. \ref{vcir} shows the resulting rotation curve, assuming a constant 
mass-to-light ratio. The figure includes the data points from the HI
observations. A constant M/L allows to fit the observations; only
in the very center, inside a radius of 300 pc (or 1.6"), we have 
reduced the M/L ratio to avoid the high luminosity peak
due to the Seyfert nucleus (but the involved mass is negligible).
The constant M/L ratio that inside a radius of 5.8 kpc give a total mass 
of 2.5 10$^{10}$ M$_\odot$. 

The gravitational potential is thus obtained in the plane, 
but we perform  the gas simulations in 3D assuming 
cylindrical symmetry for the gravitational forces within the plane.  
This hypothesis is justified considering that the gas thickness (H$_{gas}\sim$50-100pc) 
is much smaller than the stellar thickness (H$_{stars}\sim$1 kpc). 
We assume H$_{stars}$ to 
be constant with radius. The vertical z-forces are derived assuming an isothermal 
stellar disk with a sech$^2$(z/H) density law.

Once the potential is computed, we separate the axisymmetric part from
the non-axisymmetric one. The gaseous disk is set-up
initially with circular orbits in rotational
equilibrium in the axisymmetric potential. 
Progressively, with a time-scale
of 250 Myr, the non-axisymmetric part is introduced with a fixed angular speed $\Omega_p$. 
The simulations are continued until a steady response of the gas is reached, 
which appears after $\sim$ 400- 500 Myr.    
We have computed several runs,  varying the value of $\Omega_p$, until a morphology 
compatible with the observations has been obtained. 

  We have not considered here the self-gravity of the gas, since
its mass inside the optical disk is not dominant and
self-gravity will only introduce small morphological perturbations
(Sempere et al 1995). We have also adopted
a simple collisional scheme for gas clouds, without a cloud mass spectrum
(Combes \& Gerin 1985).
Clouds interact with each other via inelastic collisions,
losing 75\% of their radial relative velocity in the collision. 
The clouds initially have 
an exponential radial distribution, with a scale length of 1.5 kpc (8"). 

\subsection{ Results}

Reproducing exactly the outer spiral has not been considered as a strong constraint, 
because it is beyond the radius where
the red image is used;
yet a spiral structure forms, generated by the bar pattern, and rather
similar to the observed one.

\begin{figure}
\psfig{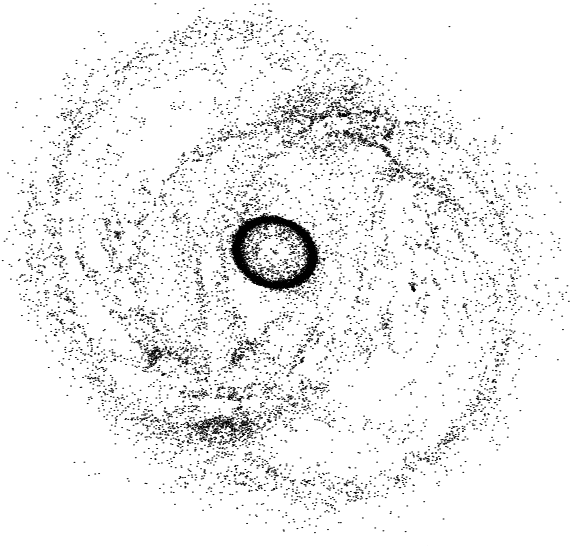}
\caption [8750.f9]{Particle plots of the $\Omega_p$ = 50 km/s/kpc run,
at steady-state.
This corresponds to a corotation at 2.5 kpc, and an outer Lindblad
resonance in the disk at 5 kpc. The particles are thus driven out to the OLR, which
does not fit the observed morphology.}
\label{om50} 
\end{figure}

\begin{figure}
\psfig{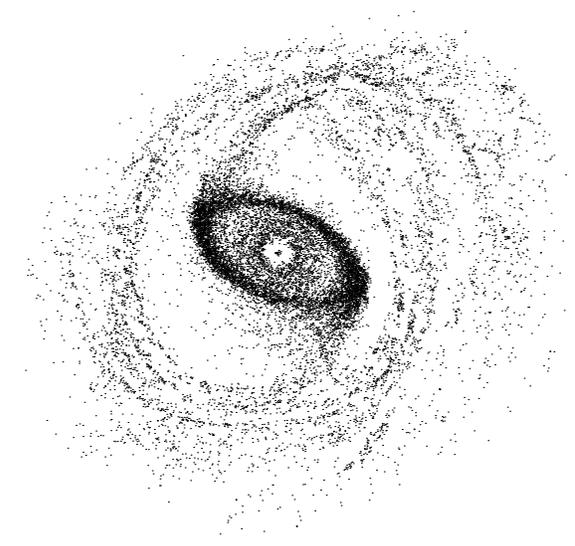}
\caption [8750.f10]{Particle plots of the $\Omega_p$ = 26 km/s/kpc run
at steady-state. The pattern speed is so low that there are two well-defined ILRs,
and the response in the region between the two ILRs
is perpendicular to the bar, which does not fit the
observations.} 
\label{om26} 
\end{figure}

\begin{figure}
\psfig{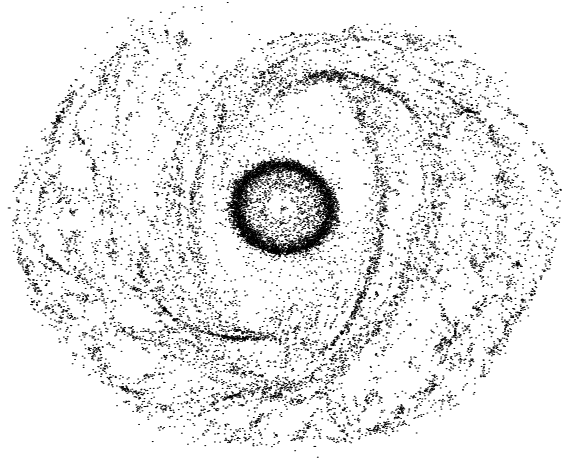}
\caption [8750.f11]{Particle plots of the run with a bar pattern speed of 
$\Omega_p$ = 38 km/s/kpc.
There is an inner ring at UHR, which corresponds to the H$\alpha$
ring at $\sim$ 3 kpc radius, but also a nuclear ring of radius 0.75 kpc.}  
\label{om38} 
\end{figure}

The simulations run with different values of $\Omega_p$,
between 20 and 60 km/s/kpc. When the pattern speed is high,
the outer Lindblad resonance moves in the visible disk, and the
matter accumulates near the OLR in a ring elongated perpendicular to the bar
(Fig. \ref{om50}).
For lower values of $\Omega_p$, the radius of corotation moves towards the outer 
parts, and there are two well separated inner Lindblad resonances.
 In between them, there is a large region where the perpendicular orbits $x_2$
dominate, generating a response perpendicular to the stellar bar
in the center, which is not observed (Fig \ref{om26}). Note that the presence of
the two ILRs cannot be inferred directly from the $\Omega - \kappa/2$ 
curve, since the bar is strong, and the potential is non-axisymmetric. 
 The best fit with observations is obtained for $\Omega_p$ = 38 km/s/kpc; the corotation
is then at 4.5 kpc, and we obtain an inner ring near the UHR, at
about the position of the observed H$\alpha$ ring.
The choice of $\Omega_p$ is therefore constrained to a narrow range 
around 38 km/s/kpc (Fig. \ref{om38}).

\subsection{ Discussion}

Angular momentum transfer is required to bring gas to the
central regions in barred galaxies (Garc\'{\i}a-Burillo et al 1993, 
1994, Sempere et al. 1995; Lindblad \& Kristen 1996; Lindblad 
et al. 1996; Sellwood \& Debattista 1996). The bar torques are able
to drive gas from the corotation radius to the center if there is no inner
Lindblad resonance, or to make the gas pile up near the ILR.
The simulations indicate that the most plausible value for 
$\Omega_p$ is 38 km/s/kpc. This implies that corotation is at about 4.5 Kpc, 
slightly beyond the end of the observed bar in the red image. 
According to the NGC 3783 light distribution, and
to its derived rotation curve, a nuclear ring near the ILR should have
been formed. This might correspond to the H$\alpha$
central concentration observed presently (see Fig. \ref{opt}).
But this accumulation should evolve quickly: either  a starburst
is triggered, that consumes the gas, or the mass concentration
is such that a secondary bar can decouple (Friedli \& Martinet 1997). 
A nested bar could take over to drive the gas to the nucleus
(Shlosman et al 1989). It is not possible, with the present spatial
resolution, to see any nuclear bar in our red image. The HST image 
obtained from the archive does not reveal either any smaller structure.
 Without any faster nuclear bar, the only possibility 
is that the gas has been driven inwards
through a small leading spiral
(e.g. Combes 1998). This is a very transient mechanism,
which would have disappeared by now.

\section{Summary}
We present the spatial distribution and kinematics of the neutral atomic 
hydrogen gas in the galaxy NGC~3783. Our main results are:

\begin{itemize}
\item Less than 2\% of the total observed HI mass is in the stellar bar
  (inside a radius of 18$''$). The total HI mass is found to be $M_{\rm HI}
  = 1.1 \times 10^9$~M$_{\odot}$. 
\item Although the HI spatial distribution appears perturbed, the HI
kinematics indicates that most of the gas is in an almost 
face on distribution ( with inclination to the line of sight 
at $i \approx 25^{\circ}$), and with the 
  kinematic major axis at $PA \approx 145^{\circ}$. 
\item The observed velocity field indicates that HI in NGC~3783 is in 
  differential rotation in a disk with a systemic velocity of $v_{\rm sys}
  \approx 2910$ km\,s$^{-1}$ and the north east region being closer to the 
  observer.
\item The rotation curve is flat up to 180$''$ from the nucleus.
\item Small velocity deviations of the order of 20 km\,s$^{-1}$ are observed
  but they may well be the result of local turbulence, since there are no
  optical signs of tidal interaction or merger event.
\item The HI spatial distribution indicates that gas can be found at large radii
  in three quadrants, NW, NE and SE. Gas in the SW quadrant is found only at
  smaller radii. 
\item If corotation occurs approximately at 1.3 times the radius of the 
  stellar bar, i.e. 24'', as suggested by numerical simulations, then the
  pattern speed of the bar is estimated to be $\Omega_{\rm b} \approx 38$
  km\,s$^{-1}$ kpc$^{-1}$ and the bright HII regions would indicate density
  enhancements near the ultra-harmonic resonance (UHR). The model also
predicts the existence of a nuclear ring at 750 pc, where an H$\alpha$
concentration is observed, indeed.
\end{itemize}

\acknowledgements{
We acknowledge helpful comments and suggestions by Tom Jones.
JAG-B acknowledges partial financial support from DGAPA-UNAM (Mexico) and 
from CONACYT (Mexico) that allowed him to spend his sabbatical year at 
the Department of Astronomy at the University of Minnesota. JAG-B would 
like to thank the hospitality of the Department of Astronomy at the University of Minnesota, 
where most of the analysis was done and the first versions of the paper was written and 
specially John Dickey for his help and useful comments about this work. The work of JF 
was partially supported by a DGAPA (UNAM) grant, CONACYT (Mexico) grants 
400354-5-4843E and 400354-5-0639PE, and by a R\&D CRAY Research grant. 
This research has made use of the NASA/IPAC extragalactic database (NED) 
which is operated by the Jet Propulsion Laboratory, Caltech, under contract
with the National Aeronautics and Space Administration. }

\end{document}